\documentclass[aps,superscriptaddress,twocolumn,amssymb]{revtex4-1}
\usepackage{revsymb4-1}
\usepackage{amsmath}
\usepackage{siunitx}
\usepackage{graphicx}
\usepackage{amssymb}
\usepackage{multirow}
\usepackage{float}
\usepackage{color}
\usepackage{comment}
\usepackage{natbib}
\usepackage{booktabs}
\usepackage{filecontents}
\usepackage{chngcntr}
\usepackage[colorlinks=true, allcolors=blue]{hyperref}
\usepackage[version=3]{mhchem} 

\newcommand{\upperRomannumeral}[1]{\uppercase\expandafter{\romannumeral#1}}

\begin{document}

\title{Two-dimensional chromium bismuthate: A room-temperature Ising ferromagnet with tunable magneto-optical response}

\author{A. Mogulkoc}
\email{mogulkoc@science.ankara.edu.tr}
\affiliation{Department of Physics, Faculty of Sciences, Ankara University, 06100 Tandogan, Ankara, Turkey}
   
\author{M. Modarresi}
\affiliation{Department of Physics, Ferdowsi University of Mashhad, Mashhad, Iran}

\author{A.~N. Rudenko}
\email{a.rudenko@science.ru.nl}
\affiliation{\mbox{Radboud University, Institute for Molecules and Materials, Heyendaalseweg 135, 6525 AJ Nijmegen, Netherlands}}
\affiliation{\mbox{Department of Theoretical Physics and Applied Mathematics, Ural Federal University, 620002 Ekaterinburg, Russia}}

\date{\today}

\begin{abstract}
We present a density functional theory (DFT) based study of a two-dimensional phase of chromium bismuthate (CrBi), previously unknown material with exceptional magnetic and magnetooptical characteristics. 
Monolayer CrBi is a ferromagnetic metal with strong spin-orbit coupling induced by the heavy bismuth atoms, resulting in a strongly anisotropic Ising-type magnetic ordering with the Curie temperature estimated to be higher than 300 K. 
The electronic structure of the system is topologically nontrivial, giving rise to a nonzero Berry curvature in the ground magnetic state, leading to the anomalous Hall effect with the conductivity plateau of $\sim$1.5 $e^2/h$ at the Fermi level. Remarkably, the Hall conductivity and the magnetooptical constant are found to be strongly dependent on the direction of magnetization. 
Besides, monolayer CrBi demonstrates the polar magnetooptical Kerr effect in the visible and near-ultraviolet spectral ranges with the maximum rotation angles of up to 10 mrad. Our findinds suggest that monolayer CrBi is a promising system for practical applications in magnetooptical and spintronic devices.

\end{abstract}

\maketitle
\section{Introduction}
The discovery of graphene and its atom-thick analogues with unique electronic properties has attracted a great interest to two-dimensional (2D) materials \cite{Wang_2020, Zavabeti_2020, Khan_2020, Bernardi_2017}. 
Recent experimental observation of magnetic order in 2D van der Waals crystals \cite{Gong_2017,Huang_2017,Chen_2019,Bonilla_2018,O_Hara_2018,Hattori_2019,Deng_2018} was encouraging from both fundamental and practical points of view.
In practice, 2D magnets offer exceptional tunability by means of applied voltage \cite{Huang_2018, Wang_2018-2, Jiang_2018,Su_rez_Morell_2019}, mechanical strain \cite{Wu_2019, Guan_2020, Lv_2019, Albaridy_2020}, stacking \cite{Huang_2017, Gong_2017, Chen_2019, Jiang_2018-2,Sivadas_2018,Soriano_2019}, 
and can be used as electrodes for prospective spintronic and magnetoresistance nanodevices \cite{Klein_2018, Song_2018, Wang_2018}. The presence of magnetic order in two dimensions is a question of fundamental importance.
Unlike three-dimensional magnets, long-range magnetic order in isotropic 2D magnets at finite temperatures is unstable with respect to thermal spin fluctuations 
\cite{mermin, Hohenberg_1967}.
The presence of an energy gap in the spectrum of spin-wave excitations is essential to stabilize the magnetic order in two dimensions, which is usually a direct consequence of magnetic anisotropy \cite{Bruno_1991,Irkhin_1999}. 

Experimental demonstration of 2D magnetism has triggered intensive theoretical research in this field. Theoretical studies are mainly focused on experimentally available materials, such as 2D
chromium trihalides \cite{Lado_2017, Webster_2018, Li_2019, Kashin2019, Kvashnin_2020, Soriano_2019, Su_rez_Morell_2019, Behera_2019,Tiwari_2021} and vanadium dichalcogenides \cite{Esters_2017, Pushkarev_2019, Luo_2017,Zhou_2019, Isaacs_2016}.
On the other hand, quite a few 2D magnets have been predicted on the basis of first-principles calculations \cite{PhysRevB.102.024441, Han_2020, Chen_2020-1, Liu_2017, Li_2020, Chen_2020, Chen_2020-2, Choudhary_2020, Haastrup_2018, Rhone_2020, Miyazato_2018,Reyntjens_2020}.
Among the large variety of 2D materials, one can distinguish materials with strong spin-orbit coupling (SOC), which 
is usually realized in heavy-element compounds. Apart from the fundamental interest, strong SOC provides additional control over 
materials' properties by means of electric \cite{Prishchenko_2018, Lugovskoi_2019} or magnetic fields \cite{Jiang_2018}, as well as plays a key role in the realization of topologically nontrivial phases \cite{Hasan_2010}.
In magnetic 2D materials, the effect of SOC 
is even more remarkable. First, it gives rise to magnetic anisotropy, which is responsible for the stabilization of long-range magnetic
order in two dimensions \cite{Lado_2017}. Second, the breaking of time-reversal symmetry in combination with SOC affects the light-matter interaction, giving rise to anomalous transport properties \cite{Nagaosa_2010, Liu_2016} and to magnetooptical effects, such as the Faraday or Kerr rotation of the polarization angle \cite{WuM,Kumar_Gudelli_2019,Molina_S_nchez_2020,Catarina_2020,KeYang}. Practically, the magnetooptical effects is a useful tool for the characterization of 2D magnetic materials. For example, magnetooptical measurements give insights into the variation of magnetic order from monolayer to multilayers \cite{Huang_2017}. On the other hand, magnetic materials with large Kerr or Faraday rotations are promising candidates for potential magnetooptic devices such as controllable optical polarizers, modulators, switches, etc. 

Recently, ferromagnetism has been predicted in 2D chromium pnictides 
\cite{Kuklin_2017, PhysRevB.102.024441, Rahman_2019}. 
This was motivating us to consider another member of this family, namely, hexagonal monolayer CrBi. 
We show that the combination of strong SOC from the heavy bismuth atoms and the magnetic coupling between Cr atoms leads to exceptional properties of this material, such as room temperature Ising-like ferromagnetism and tunable magnetooptical response. Specifically, we start from density functional theory (DFT) calculations, and investigate dynamical stability and electronic properties of monolayer CrBi. We then construct an effective anisotropic spin Hamiltonian and determine its parameters from first-principles calculations. We find that monolayer CrBi is a metallic ferromagnet with the easy axis normal to the 2D plane. Due to strong single-ion anisotropy, the system behaves like an Ising ferromagnet with the Curie temperatures around 400 K. We also find that strong SOC gives rise to a nonzero Berry curvature resulting in the anomalous Hall conductivity of $\sim$1.5 $e^2/h$. Besides, our calculations show that monolayer CrBi demonstrates the magnetooptical Kerr effect in the visible spectral range with the rotation angles up to 10 mrad, which is strongly sensitive to the direction of magnetization. 

The rest of the paper is organized as follows. Section \ref{method} briefly describes the theoretical model and computational details. In Sec.~\ref{sec3a}, we describe the atomic structure and assess the dynamical stability of monolayer CrBi. We then discuss the electronic structure (Sec.~\ref{sec3b}), and finite-temperature magnetism (Sec.~\ref{sec3c}). Section \ref{sec3d} is devoted to the anomalous transport properties and their discussion. In Sec.~~\ref{sec3e}, we analyze magnetooptical properties of monolayer CrBi. In the last section (Sec.~\ref{sec4}), we briefly summarize our results.

\section{\label{method}Method and computational details}
\subsection{First-principles calculations}
The ground-state electronic structure calculations are performed within the \emph{ab initio} methods based on DFT. We adopt the projected augmented wave (PAW) method \cite{paw,paw2} as implemented in the \emph{Vienna Ab initio Simulation Package} ({\sc vasp}) \cite{KRESSE199615,PhysRevB.54.11169}. The Brillouin zone (BZ) is sampled with a ($32\times 32\times1$) {\bf k}-point mesh by using the Monkhorst–Pack algorithm \cite{PhysRevB.13.5188}. The kinetic energy cut-off for the plane-wave basis set is set to 800 eV. The exchange and correlation effects were described by using the generalized gradient approximation with the Perdew-Burke-Ernzerhof functional \cite{PhysRevLett.77.3865}. To capture the effects of strongly correlated electrons in atomic $d$ orbitals, we use the DFT+$U$ method \cite{ldapu,PhysRevB.57.1505} applying the effective on-site Coulomb repulsion to the $d$ orbitals of Cr atoms. The value of effective Hubbard-$U$ parameter depends on the chemical structure and can be modulated by substrate dielectric screening and environment effects. Here we use $U=3$ eV, which is a typical value for Cr-based compounds and consistent with previous DFT+$U$ studies of Cr-based monolayer structures \cite{Kuklin_2017, Modarresi_2019, Lado_2017}. The lattice and atomic positions are optimized by using the conjugate gradient method to reach the ground state energy configuration with a maximum $0.001$ eV/\AA{} force tolerance on each atom. The convergence criterion for the self-consistent solution of Kohn-Sham equations is set to 10$^{-8}$ eV between two sequential steps. To avoid the interaction between periodic supercell images, we use a $30$ \AA{} vacuum distance along the non-periodic \textit{z}-direction. The dynamical and thermal stability is examined by the calculation of phonon spectra and $ab~initio$ molecular dynamics (AIMD) simulatons. The phonon spectra were obtained using the {\sc phonopy} code \cite{phono3py} which is based on density functional perturbation theory, and using a (5$\times$5$\times$1) supercell. The AIMD simulations were performed within the canonical ensemble (NVT) and Nos\'{e}-Hoover thermostat at $T=300$ K for a (4 $\times$ 4 $\times$ 1) supercell with a time step of 2 fs. The charge distribution between Cr and Bi atoms is calculated using the Bader population analysis 
\cite{0953-8984-21-8-084204,HENKELMAN2006354}.
The side and top views of the CrBi hexagonal lattice (point group $D_{3h}$) are presented in Fig.~\ref{structure}(a).

The interband optical conductivity is calculated in the independent-particle approximation with the Kubo-Greenwood formula, which reads
\begin{eqnarray}
\sigma^{inter}_{\alpha \beta}(\omega)&=&-\frac{ie^{2} \hbar}{N_{k} S }\sum_{\mathbf{k}} \sum_{n,m} \frac{f_{m\mathbf{k}}-f_{n\mathbf{k}}}{\varepsilon_{m\mathbf{k}}-\varepsilon_{n\mathbf{k}}} \\ \notag
&\times&\frac{\left\langle \psi_{n\mathbf{k}} |v_{\alpha}| \psi_{m\mathbf{k}} \right\rangle \left\langle \psi_{m\mathbf{k}} |v_{\beta}| \psi_{n\mathbf{k}} \right\rangle}{\varepsilon_{m\mathbf{k}}-\varepsilon_{n\mathbf{k}}-\left(\hbar \omega -i \eta \right)}.
\label{kubo_1}
\end{eqnarray} 
Here, $\alpha$ ($\beta$) denote the Cartesian directions, $S$ is the cell area, $N_{k}$ is the number of {\bf k}-points used for BZ sampling, $f_{n\mathbf{k}}=f(\varepsilon_{n\mathbf{k}})$ is the Fermi-Dirac distribution function, and $v_{\alpha(\beta)}$ is the $\alpha$ ($\beta$) component of the group velocity operator defined in ${\bf k}$-space. $\omega$ is the optical frequency, and $\eta = 0.01$ eV is an adjustable smearing parameter. The calculations are performed using the {\sc Berry} module \cite{PhysRevB.74.195118,PhysRevLett.92.037204} of the {\sc wannier90} package \cite{MOSTOFI2008685}, which implements an efficient band interpolation technique by means of the maximally localized Wannier functions \cite{PhysRevB.56.12847,RevModPhys.84.1419}. 

 Taking into account the metallic character of CrBi, the intraband transitions should be also considered on top of the interband transition, which can be described by the conventional Drude term. The corresponding contribution to the diagonal components of the optical conductivity tensor is given by \cite{fox2001optical},
\begin{eqnarray}
\sigma_{\alpha \alpha}^{intra}\left(\omega\right)&=&\varepsilon_{0} \frac{h\omega_{\alpha,p}^{2}}{\left(\gamma-i\omega\right)}. 
\label{kubo_3}
\end{eqnarray} 
Here, $\gamma$ is a damping term which is taken to be 0.01 eV assuming the low scattering regime, and $\omega_{\alpha,p}$ is the $\alpha$-component of the effective plasma frequency,
\begin{equation}
\omega^2_{\alpha,p}=-\frac{e^{2}}{Sh\varepsilon_{0}}\sum_{n,\bf k}\left(\frac{\partial f_{n{\bf k}}}{\partial \varepsilon_{n{\bf k}}}\right)|\langle \psi_{n{\bf k}}|v_{\alpha}|\psi_{n{\bf k}}\rangle|^2,
\end{equation}
where we have introduced an effective layer thickness $h$. We assume $h$=4.6 \AA{} for monolayer CrBi, considering the van der Waals radius of Bi atom, which is 2.3 \AA{} \cite{doi:10.1021/jp8111556}. 
In what follows, the total optical conductivity is understood as the sum of the two terms, $\sigma_{\alpha \beta}=\sigma^{intra}_{\alpha \beta}+\sigma^{inter}_{\alpha \beta}$. The corresponding complex frequency-dependent dielectric function in the long-wavelength limit reads 
\begin{eqnarray}
\varepsilon_{\alpha \beta} \left(\omega\right)&=&\delta_{\alpha\beta}+\frac{i\sigma_{\alpha \beta}\left(\omega\right)}{h\omega \varepsilon_0}.
\label{kubo_4}
\end{eqnarray}
It is convenient to define the magnetooptical constant as
\begin{equation} 
Q=\frac{i\varepsilon^{A}_{xy}}{\varepsilon_{xx}}, \label{voigt_const}
\end{equation}
where $\varepsilon^{A}_{xy}$ is the asymmetric part of the off-diagonal component of the dielectric tensor.

In the limit of zero frequency $\omega \rightarrow 0$, the asymmetric off-diagonal component of the conductivity tensor corresponds to the anomalous Hall conductivity (AHC). It can be represented in terms of the $z$-component of the Berry curvature $\Omega_{n,z}(\mathbf{k})$ as
\begin{equation} 
\sigma_{xy}(0)=-\frac{e^2}{\hbar S}\sum_{n,{\bf k}}f_{n}(\mathbf{k})\Omega_{n,z}(\mathbf{k}).
\label{AHC_eq}
\end{equation}
The Berry curvature is given by
\begin{equation} 
\Omega_{n}(\mathbf{k})=\nabla_{\mathbf{k}}\times\mathbf{A}_{n}(\mathbf{k})
\label{Berry_eq}
\end{equation}
where $\mathbf{A}_{n}(\mathbf{k})=\left\langle \psi_{n\mathbf{k}} |i\nabla_{\mathbf{k}}| \psi_{n\mathbf{k}} \right\rangle$ is the Berry connection defined in terms of cell-periodic Bloch states $|u_{n\mathbf{k}} \rangle=e^{i\mathbf{k}\cdot \mathbf{r}} | \psi_{n\mathbf{k}} \rangle$.

When a polarized light is reflected from a ferromagnetic surface, its polarization state changes, which is the hallmark of
the magnetooptical (polar) Kerr effect. The effect is characterized by the Kerr rotation ($\phi_{K}$) and ellipticity ($\psi_{K}$) angles. In the limit $h \ll c\omega^{-1}$, i.e. when the material thickness is smaller compared to the photon wavelength, the following asymptotic expression is applicable for the case of normal incidence \cite{zvezdin1997modern},
\begin{equation} 
\phi_{K}-i\psi_{K}=\frac{2nQ\omega h}{c(n^{2}-1)}.
\label{kerr}
\end{equation}
Here, $n=\sqrt{\varepsilon_{s}}$ is the refractive index of a substrate (SiO$_{2}$ $\varepsilon_{s}=2.4$ \cite{wolf1963american}), and $c$ is the speed of light.

\subsection{Anisotropic spin model}
The DFT results can be mapped onto the anisotropic classical spin model,
\begin{equation}
H=-\sum_{i,j}J_{ij} {\bf S}_{i}\cdot {\bf S}_{j} - \sum_i D_i ({\bf S}_i \cdot {\bf z}_i)^2-\sum_{i,j}\delta_{ij} S_{i}^{z}S_{j}^{z},
\label{heis}
\end{equation}
where $J_{ij}\equiv J({\bf R}_{ij})$ is the Heisenberg exchange interaction between spins ${\bf S}_{i}$ and ${\bf S}_{j}$, $D_{i}$ is the single-ion magnetic anisotropy at spin ${\bf S}_i$, $\delta_{ij}$ is the anisotropic exchange interaction
, and ${\bf z}_i$ is the unit vector pointing in the direction of the easy magnetization axis. 
Strictly speaking, Eq.~(\ref{heis}) is justified for the atomic limit, i.e. when the Coulomb interaction between the Cr 3$d$ electrons is larger compared to the 3$d$ band width. We assume that this condition is fairly fulfilled in CrBi considering that the magnetization density is well-localized on Cr atoms, leading to the integer magnetic moments. Nevertheless, we note that spin excitations resulting from Eq.~(\ref{heis}) should be considered as approximate, which might lead, for instance, to overestimation of the critical temperatures.

For simplicity, we consider only nearest-neighbor exchange interactions in Eq.(\ref{heis}). 
The Mermin-Wagner-Hohenberg theorem states the absence of finite temperature phase transition for the isotropic Heisenberg Hamiltonian with short range interactions \cite{mermin,Hohenberg_1967}. As a consequence, the anisotropy energy is crucially important for the stabilization of magnetic order in two dimensions \cite{Modarresi_2019,Lado_2017,Li_2019}. In the hexagonal CrBi lattice each Cr atom has six nearest neighbor Cr atoms. According to this model, the exchange coupling parameters can be expressed as \cite{supplement},  
\begin{eqnarray}
\label{param}
J&=&\left(E_{AFM_{x}}-E_{FM_{x}}\right)/16S^{2},  \\
\notag
D&=&\left(E_{FM_{x}}+3E_{AFM_{x}}-E_{FM_{z}}-3E_{AFM_{z}}\right)/4S^{2}, \\
\notag
\delta&=&\left(E_{FM_{x}}+E_{AFM_{z}}-E_{FM_{z}}-E_{AFM_{x}}\right)/16S^{2}
\end{eqnarray}
Here, $E_{FM_x}$, $E_{FM_z}$, $E_{AFM_x}$, and $E_{AFM_z}$ are the total energies of FM$_x$, FM$_z$, AFM$_x$ and AFM$_{z}$ configurations, respectively, shown in Fig.~\ref{structure}(b).
The ground state of CrBi corresponds to a FM order with the magnetic moments $M=3.0$ $\mu_{B}$ localized on Cr atoms, which corresponds to $S=3/2$. The spin operators in the Hamiltonian Eq.~(\ref{heis}) can be transformed into bosonic variables by means of linear spin-wave theory. The resulting spin-wave spectrum at zero temperature reads \cite{VSe2_paper}.
\begin{equation} 
\omega({\bf q})=2S(D+z\delta+J(0)-J({\bf q})),
\end{equation}
where $J({\bf q})$ is the Fourier transform of $J({\bf R}_{ij})$ and $z=6$ is the number of nearest neighbors. The spin-wave spectrum has a gap $\Delta=2S(D+z\delta)$, which stabilizes the 2D magnetic order. In the limit of a vanishing gap ($\Delta \rightarrow 0$), the long-range order at finite temperatures disappears, i.e. the ordering temperature $T_C  \sim -1/\mathrm{ln}(\Delta) \rightarrow 0$ \cite{Modarresi_2019}, in accordance with the Mermin-Wagner-Hohenberg theorem. In the presence of a dielectric substrate, the on-site Coulomb interaction on Cr atoms could be screened leading to smaller $U$ values. In order to understand the role of screening, $J$, $\delta$ and $D$ are calculated as a function of the Hubbard \textit{U} parameter and presented in Fig. S2 \cite{supplement}. It is known that next nearest-neighbor exchange interactions can affect the Curie temperature  in 2D magnets \cite{Vanherck_2020}. To check the effect of second nearest-neighbor exchange interaction in CrBi, we map the DFT results onto the Heisenberg Hamiltonian with $J_1$ and $J_2$ presented in Supplemental Material \cite{supplement}. 


\section{Result and discussion}
\subsection{\label{sec3a}Dynamical and thermal stability}

\begin{figure*}[b]
\includegraphics[width=0.95\linewidth]{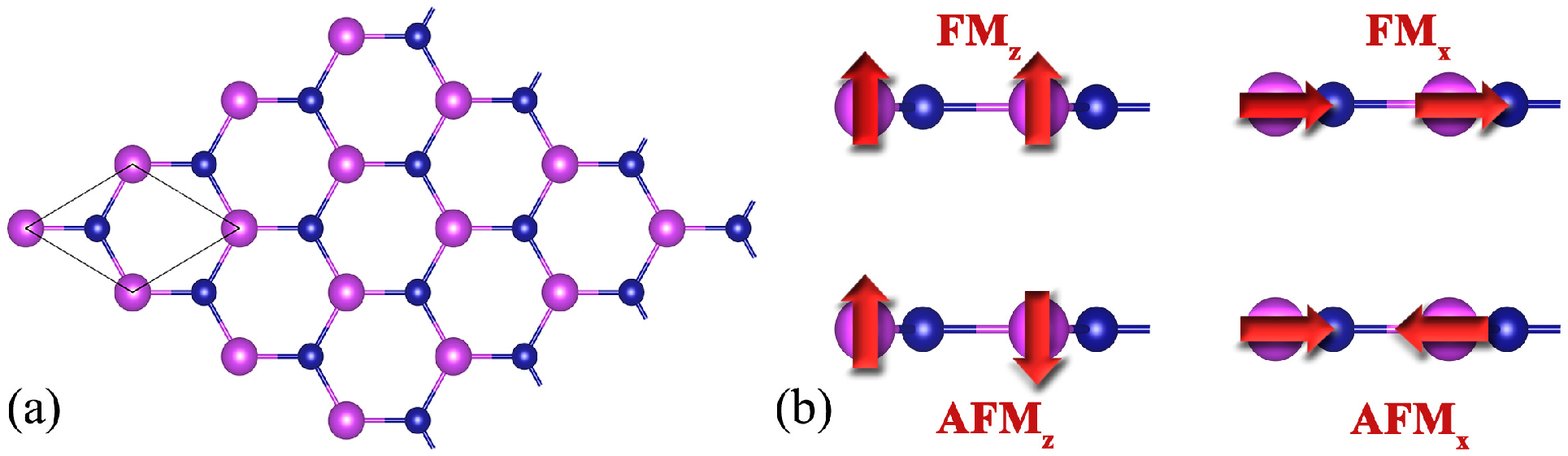}
\caption{The (a) top and (b) side view of relaxed monolayer CrBi together with the FM and AFM configurations along the \textit{x} and \textit{z} directions. The purple and blue colors correspond to Cr and Bi atoms, respectively.
}
\label{structure}
\end{figure*}


\begin{figure*}[b]
\includegraphics[width=\linewidth]{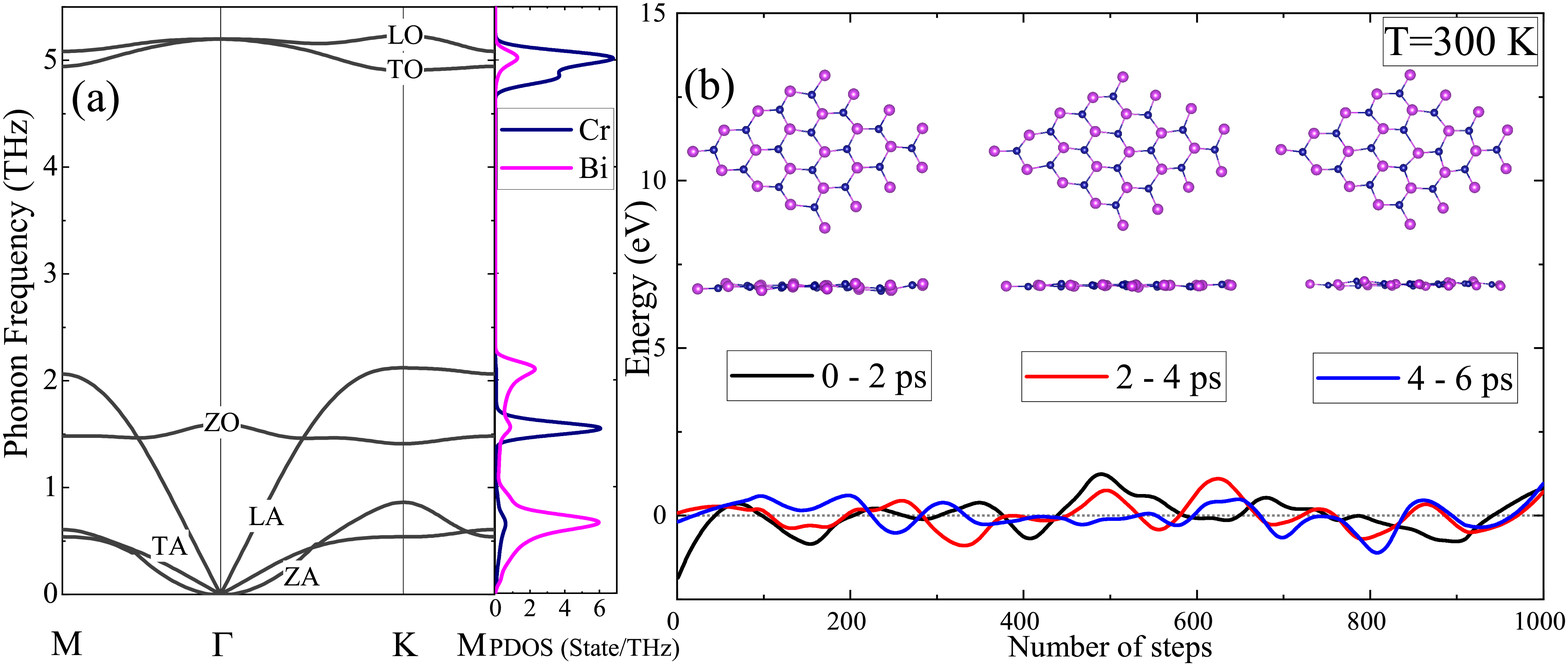}
\caption{(a) Phonon dispersion together with partial density of states (PDOS) calculated for monolayer CrBi with $U=3$ eV.  (b) Total energy obtained from the AIMD simulation performed over a period of 6 ps with NVT ensemble at $T=300$ K. }
\label{phonon}
\end{figure*}



We first examine the dynamical stability of the CrBi crystal structure by calculating the phonon spectrum presented in Fig.~\ref{phonon}(a). There are three acoustic and three optical phonon branches. Around the $\Gamma$ point the acoustic phonons are represented by two in-plane modes (LA and TA) with linear dispersion $\omega\sim |{\bf k}|$, and one flexural mode (ZA) with quadratic dispersion $\omega\sim {\bf k}^2$, which is typical to 2D materials \cite{PhysRevB.100.075417}. The characteristic frequency of acoustic phonons, given by the freuencies at the edges of BZ, appears at $\sim$0.6 THz, and originates predominantly from the vibrations of Bi atoms [see partial density of states in Fig.~\ref{phonon}(a)]. In contrast, the optical phonons are characterized by essentially dispersionless modes mainly originating from Cr vibrations. There is one out-of-plane (ZO) and two in-plane (LO and TO) modes appearing around 1.5 and 5 THz, respectively. One can see the absence of imaginary modes in the entire BZ, which indicates stability of the crystal structure with respect to atomic vibrations. 
It is worth noting that the system may lose its stability in the presence of highly dielectric environment (e.g., substrate), as demonstrated by our calculations performed without the Hubbard $U$ correction (not presented here).

We also analyze 
thermal stability of monolayer CrBi using the AIMD simulations carried out with NVT ensemble at room temperature for a time period of 6 ps. The corresponding fluctuations of the total energy, as well as the structure snapshots taken every 2 ps are presented in Fig.~ \ref{phonon}(b). 
One can see that the atomic structure remains stable and essentially flat at $T=300$ K. Some distortions from the perfect hexagonal lattice are not large, and can be related to thermal fluctuations. The calculated root mean square displacement averaged over the last 1 ps results in $\sqrt{\langle |{\bf r}(t)-{\bf r}(0)|^2 \rangle}\approx 0.21$ \AA{}, whereas the averaged total energy fluctuation is around 30 meV/atom.  
This allows us to confirm thermal stability of monolayer CrBi at room temperature.

The structural parameters, cohesive energy, charge transfer, and work function of monolayer CrBi are summarized in Table~\ref{table1}. The monolayer CrBi has no buckling with the lattice constant of 4.80 \AA{}. The Bader population analysis \cite{HENKELMAN2006354} shows that there is a charge transfer from Cr to Bi around 0.6$e$ which suggests a considerable ionic contribution to the bonding.
The cohesive energy per unitcell of monolayer CrBi is around 3 eV, which is comparable with other monolayer chromium pnictides \cite{PhysRevB.102.024441}. To characterize the photoelectric threshold, the work function is presented in Table~\ref{table1}.
The work function of CrBi 
is considerably smaller than in other 2D ferromagnets, such as chromium trihalides and vanadium dichalcogenides, where it is around 6 eV \cite{C7NR06473J,LIU2018419, Li_2019-2}. Thus, monolayer CrBi can be a good candidate for electrocatalysis applications, offering low energy barrier for the charge transfer. On the other hand, this may affect stability of the crystal structure in the presence of chemically reactive species.

\begin{table}
 \centering
 \caption{\label{table1} 
 Lattice constant ($a$), Cr-X bond length ($\ell_{Cr-X} $), charge transfer ($\Delta Q$), cohesive energy ($E_{coh}$), and work function ($\Phi_{w}$). The calculations are performed with $U=3$ eV. The work function is defined as $\Phi_{w}=E^{vac}-E_{F}$, where $E^{vac}$ is the vacuum electrostatic potential, and $E_F$ is the Fermi energy. }  

 \begin{tabular}{lcccccccccc} 
 \hline\hline
   Structure& \textit{a} (\AA) & $\ell_{Cr-Bi} $ (\AA)& $\Delta Q$ ($\mid e \mid$)& $E_{coh}$ (eV) & $\Phi_{w}$ (eV) \\ 
 \hline
CrBi & 4.80 &2.77 & 0.57 &3.11 & 3.13\\
 \hline\hline
 \end{tabular}
 \end{table}

\begin{figure*}[!t]
\includegraphics[width=0.95\linewidth]{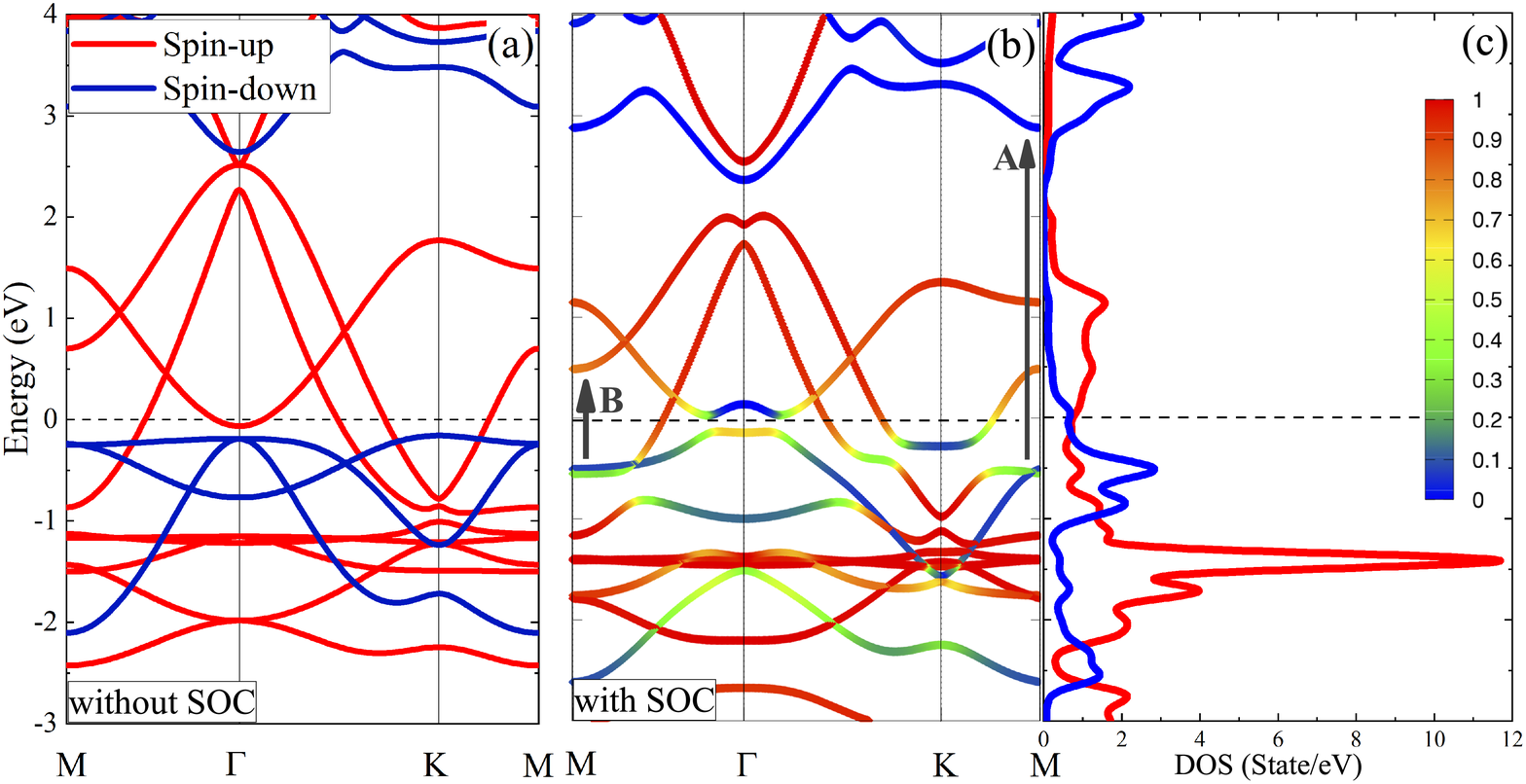}
\caption{Band structure of monolayer CrBi calculated (a) without and (b) with SOC. (c) The spin-projected DOS in the presence of SOC. Blue and red color shows the contribution of the spin-up and spin-down states. In both cases the direction of magnetization is normal to the monolayer plane (${\bf M}
\parallel{\bf z}$). The arrows indicate the most intensive optical transitions, giving rise to two prominent absorption peaks shown in Fig.~\ref{KUBO}(c).}
\label{bands}
\end{figure*}
\subsection{\label{sec3b}Electronic structure}
The electronic band structure of ferromagnetic monolayer CrBi calculated without and with SOC for the case when the magnetization is normal to the monolayer plane (${\bf M}\parallel{\bf z}$) is presented in Figs.~ \ref{bands}(a) and 
\ref{bands}(b), respectively. In the absence of SOC, monolayer CrBi is a half-metal similar to other Cr$X$ ($X$=P,As,Sb) compounds \cite{PhysRevB.102.024441}. The inclusion of SOC mixes spin-up and spin-down states, which is indicated by color of the electronic bands in Fig.~ \ref{bands}(b). Interestingly, the strongest effect of SOC is observed around the Fermi level, where one can see the largest mixing of spin-up and spin-down states. This results in a considerable modification of the electronic structure. Particularly, from Fig.~\ref{bands}(b) one can see that the band inversion is taking place accompanied by the formation of a local SOC gap of $150$ meV at avoided crossings. Remarkably, the Fermi energy lies within the gap, which suggests topologically nontrivial character of the corresponding states. Although globally the systems remains a trivial metal due to the presence of other bands crossing the Fermi energy, the SOC gap is expected to provide a considerable contribution to the anomalous transport properties.
The spin decomposed density of electronic states for the SOC case is presented in Fig.~\ref{bands}(c). 
At the Fermi level, both spin-up and spin-down states contribute equally, which is in strong contrast with the half-metallic behavior observed without SOC.
In other words, the half-metal to metal phase transition in monolayer CrBi is a manifestation of the strong SOC effect. 

The situation when the magnetization is parallel to the monolayer plane (${\bf M}\perp {\bf z}$) is different. Although the electronic bands are very similar for the $z$-axis [Fig.~\ref{bands}(b)] and $x$-axis (see Supplemental Material) magnetization, there is an important difference at the Fermi level. Specifically, in the ${\bf M}\parallel {\bf z}$ case there is a SOC-induced local gap between the valence and conduction bands at the $\Gamma$ point, while it is absent when ${\bf M}\perp {\bf z}$, and both bands are occupied. As it will be discussed below in Sec.~\ref{sec3d}, the presence of the corresponding gap produces a net Berry curvature around the $\Gamma$ point.

\subsection{\label{sec3c}Finite-temperature magnetism}

\begin{table}
 \centering
 \caption{\label{table2} 
 Isotropic exchange interaction between nearest neighbor spins ($J$), single-ion anisotropy energy ($D$), intersite anisotropy parameter ($\delta$), net magnetization per Cr atom ($M$), and the Curie temperature ($T_C$). The calculations are performed with $U=3$ eV.}  

 \begin{tabular}{lccccccccccc} 
 \hline\hline
   Parameters& $J$ (meV) & $D$ (meV) & $\delta$ (meV) & $M$ ($\mu_{B}$) & $T_C$ (K) \\ 
 \hline
CrBi & 1.59 &8.51 & 0.63 &3.0 & 420\footnote{Ising model} (600\footnote{RPA scheme}) \\
 \hline\hline
 \end{tabular}
 \end{table}

As a next step, we examine the magnetic properties of monolayer CrBi at the level of the anisotropic spin Hamiltonian given by Eq.~(\ref{heis}). In order to determine the model parameters, we estimate the total energies of different magnetic configurations shown in Fig.~\ref{structure}(b), using noncollinear DFT+$U$+SOC calculations.
The CrBi monolayer is found to have a FM ground state with the easy-axis normal to the plane. The magnetic moment is well-localized on Cr atoms, and has the value of 3.0 $\mu_B$ per cell, which corresponds to the spin $S=3/2$. The parameters $J$, $D$, and $\delta$ are determined via Eq.~(\ref{param}), where we restricted ourselves to the nearest-neighbor exchange interactions. Despite the metallic character of CrBi, the next-nearest-neighbor interactions are smaller \cite{supplement}, not affecting the results presented below. The resulting calculated parameters are 
$J=1.59$ meV, $D=8.51$ meV and $\delta=0.63$ meV, summarized in Table~\ref{table2}.
In contrast to the isostructural compounds Cr$X$ ($X$=P,As,Sb) \cite{PhysRevB.102.024441}, the isotropic exchange interaction $J$ in CrBi is considerably smaller. This can be attributed to a larger lattice constant, which suppresses FM contribution to the exchange interaction, according to the model proposed in Ref.~\cite{Kashin2019}. On the other hand, the single-ion anisotropy $D$ is significantly larger, which is due to the strong SOC of bismuth. One can see that $D \gg \delta$, i.e. the magnetic anisotropy mainly arises from the single-ion term. Although both parameters are related to SOC, they have distinct physical origin. While the anisotropic exchange $\delta$ is of kinetic origin \cite{Aharony1995}, single-ion anisotropy $D$ can be attributed to the formation of local orbital moments \cite{Bruno1989}.
Furthermore, one can see that $J \ll D$, which suggests that monolayer CrBi is an Ising-type ferromagnet. Indeed, in this regime spin fluctuations with nonzero $S_x$ and $S_y$ projections are highly unfavorable. Therefore, we can recast the spin Hamiltonian as $H\approx-\sum_{\langle ij\rangle}J^{\mathrm{eff}}_{ij}S^z_iS^z_j$, which is the Ising Hamiltonian with nearest-neighbor interactions $J^{\mathrm{eff}}_{ij}=2(J_{ij}+\delta_{ij})$. We note that the variation of the parameters with the Hubbard \textit{U} does not change the Ising-type behavior of monolayer CrBi \cite{supplement}.  

The critical temperature of the 2D Ising model on a triangular lattice is known exactly, which is   $k_BT_C = 4J^{\mathrm{eff}}S^2/\mathrm{ln}3$ \cite{Torelli_2018}. It is worth noting that the Ising model is intrinsically anisotropic, and thus compatible with the Mermin-Wagner-Hohenberg theorem. In our case the Ising model yields $T_C\simeq 420$ K. This value is expected to be somewhat overestimated because of finite $D$, and because of the ignored next-nearest-neighbor interactions, which are small but tend to destabilize the FM ordering \cite{PhysRevB.102.024441}. Nevertheless, we expect $T_C$ of the order of room temperature, which is larger than for recently proposed 2D Ising ferromagnet VI$_3$ \cite{Yang_2020} with $T_C\sim 100$ K. The high $T_C$ in CrBi is not only related to a strong exchange coupling between Cr spins, but can be also attributed to the relatively large spin $S=3/2$ and to a larger coordination number ($z=6$) compared to materials with the honeycomb lattice. It is interesting to note that a $T_C$ estimation within the RPA approach \cite{PhysRevB.71.174408,PhysRevB.102.024441} based on the Tyablikov's decoupling approximation yields even larger $T_C\simeq 600$ K (see Supplemental Material \cite{supplement} for details). However, this scheme is apparently inapplicable in the regime $D \gg J$, as demonstrated in Ref.~\cite{Torelli_2018}. The inclusion of the second-nearest-neighbor exchange interaction in the spin model does not lead to any significant changes of the Curie temperature, as we show in Supplemental Material \cite{supplement}.

\subsection{\label{sec3d}Anomalous Hall conductivity}

\begin{figure*}[!t]
\includegraphics[width=0.95\linewidth]{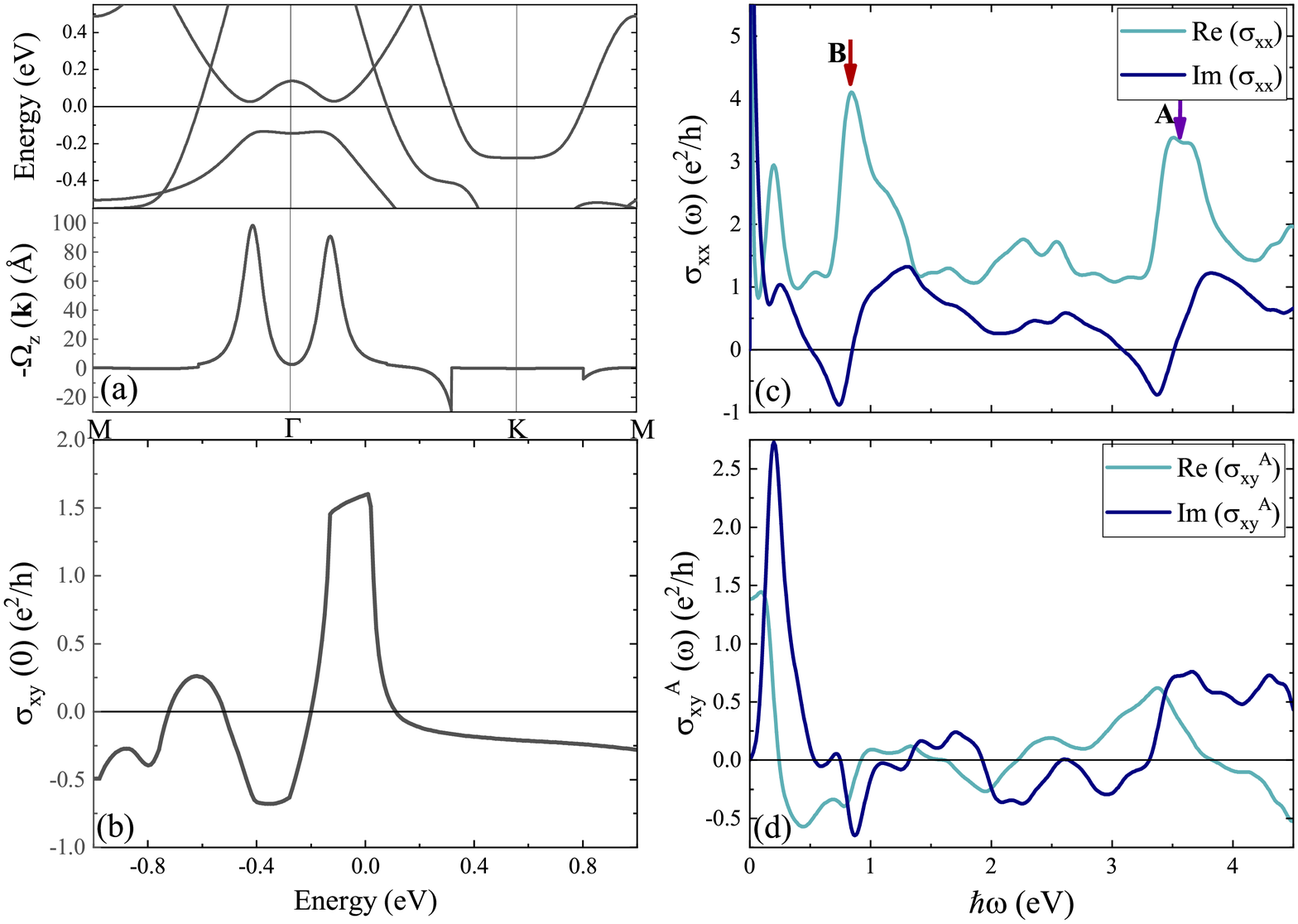}
\caption{(a) The electronic bands around the Fermi level and the Berry curvature along the high-symmetry points in BZ. (b) The anomalous Hall conductivity of monolayer CrBi calculated as a function of the energy (chemical potential). (c) Diagonal and (d) asymmetric off-diagonal components of the frequency-dependent optical conductivity. In all cases the out-of-plane magnetization ($\mathbf{M}\parallel\mathbf{z}$) of monolayer CrBi is assumed.}
\label{KUBO}
\end{figure*}

For a crystal with inversion symmetry the Berry curvature obeys $\Omega_{n}\left({\bf k}\right)=\Omega_{n}\left(-{\bf k}\right)$, while time-reversal symmetry implies $\Omega_{n}\left({\bf k}\right)=-\Omega_{n}\left(-{\bf k}\right)$, which means $\Omega_{n}\left({\bf k}\right)=0$ for a crystal with both inversion and time-reversal symmetries. In case of a magnetic monolayer, the Berry curvature may be non-zero, which is related to spontaneously broken time-reversal symmetry. Additionally, spin mixing is required to ensure a nonzero Berry curvature, which is induced by SOC. All these conditions are fullfilled in monolayer CrBi. 

Figure \ref{KUBO}(a) shows the low energy electronic bands in monolayer CrBi. Around the $\Gamma$ point the bands are inverted with a local SOC gap. At the points, where the band crossing is avoided, one can see the emergence of two strong peaks of the same sign in the Berry curvature [lower panel in Fig.\ref{KUBO}(a)]. Interestingly, if CrBi is magnetized along the in-plane ($x$) direction, the band topology around the Fermi energy changes such that the contribution to the Berry curvature around $\Gamma$ vanishes \cite{supplement}.
In Fig.~\ref{KUBO}(b), we show the anomalous Hall conductivity, which exhibits a plateau equal to $\sim$1.5 $e^{2}/h$ at the Fermi level. The width of the plateau is about 0.2 eV, which makes it robust against perturbations such as structural and charge disorder, and allows for experimental detection. The AHC plateau at the Fermi level is equivalent to $\sim$200 S/cm, which is comparable to experimental measurements for Heusler alloys \cite{Husmann_2006}, layered iron doped TaS$_2$ \cite{Checkelsky_2008}, but almost an order of magnitude smaller than in pure Fe, Ni, Co and Gd films \cite{Miyasato_2007}. It should be emphasized that in our case AHC is not quantized because the system remains a trivial metal even in the presence of SOC. The integer (quantum) AHC was theoretically reported in CrI$_3$ \cite{Zhu_2020-2}, iron trihalides \cite{Li_2019}, and CoBr$_2$ \cite{Chen_2017} monolayers.


In Fig.~\ref{KUBO}(b), one can see another AHC plateau of around $-0.5$ $e^{2}/h$ at $\sim$0.3 eV below the Fermi energy. Interestengly, AHC switches its sign for the chemical potentials around $-0.2$ and 0.1 eV. In practice, the variation of the chemical potential can be achieved by the electron or hole doping. Strong sensitivity of AHC to moderate charge doping can be regarded as a peculiarity of monolayer CrBi, and can be utilized to verify our findings experimentally. 
Another prominent characteristic of AHC in monolayer CrBi is its dependence of the magnetization direction. 
Although the $z$-direction of magnetization is the ground state of monolayer CrBi direction, the magnetic moments can be aligned in-plane upon application of external magnetic field. 
In this situation, AHC is almost zero for all practically relevant chemical potentials \cite{supplement}. 
These results are promising for the realization of \emph{tunable} anomalous Hall effect by means of external magnetic field.

\subsection{\label{sec3e}Magnetooptical response}
\begin{figure*}[!t]
\includegraphics[width=\linewidth]{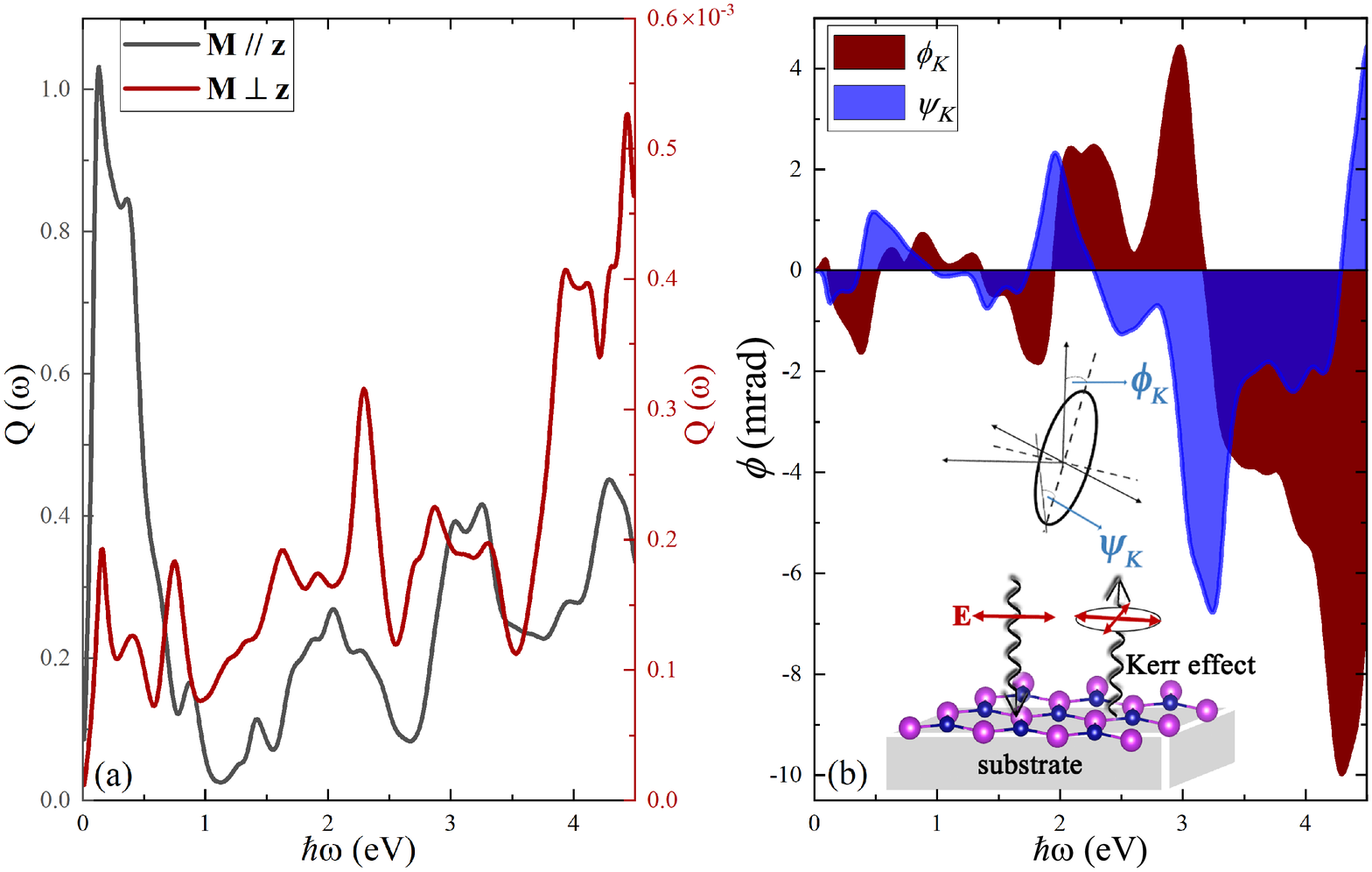}
\caption{(a) The magnetooptical constant as a function of the photon energy for the out-of-plane (${\bf M}\parallel{\bf z}$) and in-plane (${\bf M}\perp{\bf z}$) magnetization (note different scale). (b) The Kerr rotation ($\phi_K$) (red) and the Kerr ellipticity ($\psi_K$) (blue) angles calculated as a function of the photon energy for the out-of-plane (${\bf M}\parallel{\bf z}$) magnetization. In the calculations, we assume a SiO$_{2}$ substrate with the dielectric constant $\varepsilon_{s}=2.4$. Inset: Schematic illustration of the polar magnetooptical Kerr effect at normal incidence.}
\label{MO}
\end{figure*}
The real and imaginary parts of the diagonal component of the optical conductivity $\sigma_{xx}(\omega)$ are shown in Fig.~\ref{KUBO} (c) as a function of the photon energy. Due to the metallic character of CrBi, one can see a Drude peak appearing in the low frequency range originating from 
the intraband transitions included in the form of 
Eq.~(\ref{kubo_3}). 
The optical spectrum exhibits two prominent absorption peaks around $\sim$0.8 eV and $\sim$3.5 eV placed at mid-infrared (mid-IR) and near-ultraviolet (near-UV) regions, which can be directly related to the interband transitions. The corresponding transitions are shown in Fig.~\ref{bands} (b) denoted by A and B letters.
Figure \ref{KUBO}(d) shows  frequency dependence of the asymmetric off-diagonal component of the conductivity, $\sigma^{A}_{xy}(\omega)$. A nonzero $\sigma^{A}_{xy}$ is conditioned by the absence of time-reversal symmetry and the presence of SOC.
At $\sim$0.25 eV the real part has a strong peak, which can be associated with the SOC-induced gap in the band structure [Fig.~\ref{bands}(b)], and with the spin mixture of the corresponding bands at the Fermi energy. Importantly, this peak is absent in case of in-plane magnetization (${\bf M}\perp{\bf z}$) because the corresponding spin-mixed states are fully occupied (see Supplemental Material \cite{supplement}).

The magnetooptical (Voigt) constant $Q(\omega)$ arising from the off-diagonal terms of dielectric tensor [see Eq.~(\ref{voigt_const})] gives information about the magnetooptical activity. In Fig.~\ref{MO}(a), we present $Q(\omega)$ calculated for two directions of magnetization in monolayer CrBi. For the out-of-plane (${\bf M}\parallel{\bf z}$) magnetization, $Q(\omega)$ exhibits a strong peak around in the IR region similar to that observed in Fig.~\ref{KUBO}(d). On the other hand, for the in-plane (${\bf M}\perp{\bf z}$) magnetization $Q(\omega)$ is almost $\sim10^{3}$ times smaller. A similar trend is observed in the visible part of the spectrum. 
Therefore, the magnetooptical activity of monolayer CrBi is strongly dependent on the direction of magnetization, and almost negligible in the case of in-plane magnetization. We note that this is a direct consequence of the magnetization-dependent electronic structure modification, and not related to the angle of incidence.

The Kerr rotation angles are calculated by using Eq.~(\ref{kerr}) and presented as a function of the photon energy in Fig.~\ref{MO}(b) for the out-of-plane magnetization (${\bf M}\parallel{\bf z}$) and normal incidence. 
In our calculations we assume the presence of a SiO$_2$ substrate with the dielectric constant $\varepsilon_s=2.4$. As one can see from Fig.~\ref{MO}(b), for the out-of-plane magnetization (red shaded curve) that the Kerr rotation spectra ($\phi_K$) exhibits a few peaks
in the visible part of the electromagnetic spectrum with the magnitude up to 4.4 mrad ($\ang{0.25}$). The same order of rotation angles can be seen in the Kerr ellipticity spectrum ($\psi_K$). 
The Kerr rotation and ellipticity spectra both have a prominent peak in the near-UV region with magnitudes of $-10$ mrad ($\ang{-0.57}$) and $-6.8$ mrad ($\ang{-0.38}$), respectively. The calculated rotation angles are smaller, but comparable with those in ferromagnetic monolayer CrI$_3$ \cite{Kumar_Gudelli_2019,WuM,Molina_S_nchez_2020}, where the magnetooptical Kerr effect is pronounced. We note that the magnitude of $\phi_K$ and $\psi_K$ could be increased by a factor $\varepsilon_s-1$ in the free-standing monolayer.


It should be noted that in our calculations we ignored nonlocal correlation effects, responsible for the formation of bound electron-hole states. These effects are important for insulator and semiconductors because they determine the optical gap. Futhermore, reduced dielectric screening of 2D materials usually enhances the binding energies of excitons. However, in our case we deal with a metallic system, implying that the screening effects are large. In this situation, we do not expect that many-body perturbative corrections (e.g., at the $GW$ level) would significantly modify our results. 
Nevertheless, we admit that a more sophisticated many-body treatment of electronic correlations would be desirable to support our findings. This problem is left for future research.

\section{\label{sec4}Conclusion}
In summary, we have systematically studied the electronic, magnetic and magnetooptical properties of hexagonal monolayer CrBi using first-principles calculations. As a first step, we demonstrate that the crystal structure of monolayer CrBi is dynamically stable at room temperature. We then show that due to the heavy bismuth atoms, the system exhibits strong SOC, giving rise to a strong single-ion anisotropy. This allows us to consider monolayer CrBi as a 2D Ising ferromagnet, whose Curie temperature is estimated to be above 300~K. From the electronic structure of view, monolayer CrBi is a metal. Strong SOC affects topology of the electronic bands, leading to a nonzero Berry curvature. This results in the anomalous Hall effect with AHC of $\sim$1.5 $e^2/h$ at the Fermi energy. Finally, we consider magnetooptical response of monolayer CrBi and find considerable Kerr rotation angles of up to 10 mrad in the visible and near-UV spectral ranges. The magnetooptical response as well as AHC can be effectively tuned by changing the magnetization direction and, therefore, make monolayer CrBi a promising candidate for practical applications. We believe that our findings could be verified experimentally, and will be useful for future theoretical and experimental studies of 2D magnetic materials.




\section*{Acknowledgments}
Authors acknowledge the Ankara University for high performance computing facility through the AYP under Grant No. 17A0443001. This work was supported by the Scientific and Technological Research Council of Turkey (TUBITAK) under Project No 119F361. The numerical calculations reported in this paper were partially performed at TUBITAK ULAKBIM, High Performance and Grid Computing Center (TRUBA resources).
   
\section*{References}
\nocite{apsrev41Control}
\bibliographystyle{apsrev4-1}
\bibliography{references}

\end{document}